\documentclass[%
 reprint,
 prl,
 amsmath,amssymb,
 aps,
]{revtex4-2}

\usepackage{graphicx}
\usepackage{dcolumn}
\usepackage{bm}
\usepackage{hyperref}
\usepackage{breakurl}

\begin{document}

\preprint{APS/123-QED}

\title{Possible coexistence of kinetic Alfv\'en and ion Bernstein modes in sub-ion scale compressive turbulence in the solar wind}

\author{Owen Wyn Roberts}
 \email{owen.roberts@oeaw.ac.at}
\affiliation{Space Research Institute, Austrian Academy of Sciences, Schmiedlstrasse 6, 8042 Graz, Austria
}
 \author{Daniel Verscharen}
\affiliation{Mullard Space Science Laboratory, University College London, Dorking RH5 6NT, UK and\\
Space Science Center, University of New Hampshire, 8 College Road, Durham, NH 03824, USA.}

\author{Yasuhito Narita}%
\affiliation{
Space Research Institute, Austrian Academy of Sciences, Schmiedlstrasse 6,8042 Graz, Austria
}
\author{Rumi Nakamura}
\affiliation{
Space Research Institute, Austrian Academy of Sciences, Schmiedlstrasse 6,8042 Graz, Austria
}

\author{Zolt\'an V\"or\"os}
\affiliation{
Space Research Institute, Austrian Academy of Sciences, Schmiedlstrasse 6,8042 Graz, Austria and \\
Geodetic and Geophysical Institute, Research Centre for Astronomy and Earth Sciences (RCAES), Sopron, Hungary}

\author{Ferdinand Plaschke}
\affiliation{
Space Research Institute, Austrian Academy of Sciences, Schmiedlstrasse 6,8042 Graz, Austria}


\date{\today}

\begin{abstract}
We investigate compressive turbulence at sub-ion scales with measurements from the Magnetospheric MultiScale Mission. The tetrahedral configuration and high time resolution density data obtained by calibrating spacecraft potential allow an investigation of the turbulent density fluctuations in the solar wind and their three-dimensional structure in the sub-ion range. The wave-vector associated with the highest energy density at each spacecraft frequency is obtained by application of the Multi-point signal resonator technique to the four-point density data. The fluctuations show a strong wave-vector anisotropy $k_{\perp}\gg k_{\parallel}$ where the parallel and perpendicular symbols are with respect to the mean magnetic field direction. The plasma frame frequencies show two populations, one below the proton cyclotron frequency $\omega<\Omega_{ci}$ consistent with kinetic Alfv\'en wave (KAW) turbulence. The second component has higher frequencies $\omega > \Omega_{ci}$ consistent with ion Bernstein wave (IBW) turbulence. Alternatively, these fluctuations may constitute KAWs that have undergone multiple wave-wave interactions causing a broadening in the plasma frame frequencies. The scale-dependent kurtosis in this wave-vector region shows a reduction in intermittency at the small scales which can also be explained by the presence of wave activity. Our results suggest that small-scale turbulence exhibits linear-wave properties of kinetic Alfv\'en and possibly ion-Bernstein/magnetosonic waves. Based on our results, we speculate that these waves may play a role in describing the observed reduction in intermittency at sub ion scales.

\end{abstract}

\maketitle


\emph{Introduction:} Plasma turbulence is an inherently multi-scale, three-dimensional phenomenon \citep{Bruno2013,Kiyani2015,Verscharen2019}. The solar wind is an easily accessible turbulent plasma, and its temperature decreases slower with heliocentric distance than expected for an adiabatically expanding gas \cite{Williams1995}. The dissipation of plasma turbulence is likely to play a role in the observed plasma heating of the solar wind. It is still a matter of ongoing research as to what degree the turbulent fluctuations are wave-like or rather characteristic of intermittent coherent structures such as vortices or current sheets.   

In the solar wind, turbulent fluctuations exist from $\sim 10^{6}$ km (termed the correlation scale, the size of the largest eddie \citep{Kiyani2015}) to electron scales at $\sim 1$km. At scales smaller than the correlation scale fluid-like nonlinear interactions between fluctuations cascade energy to smaller scales and the power spectral density (PSD) of magnetic and density fluctuations at 1a.u. have a Kolmogorov like spectral index of -5/3 \citep{Bruno2013}. In this region, the magnetic fluctuations are dominated by the transverse components. However, there is a non-negligible component of energy in compressive fluctuations \cite{Roberts2017,Verscharen2017}. At scales near the proton characteristic scales, a break is observed in the spectrum of magnetic field fluctuations \citep{Bruno2014}, and a steepening to a spectral index of $\sim-2.6$ \citep{Smith2006a}. This region is termed the sub-ion range, where plasma waves are dispersive or dissipation heats the plasma. In this range, the relative power of the compressive component of the magnetic-field fluctuations increases, eventually becoming comparable to the transverse components \cite{Kiyani2009a}. 

 A variety of linear waves can exist in a plasma, and when their frequencies reach the characteristic scales of the particles then wave-particle interactions can occur, transferring energy from the wave to the plasma \cite{Chen2019}. At ion inertial scales the turbulence is thought to contain both kinetic Alfv\'en waves, and slow waves \cite{Roberts2017,Safrankova2019}. At smaller scales, the kinetic slow waves are damped strongly leaving only kinetic Alfv\'en waves which become increasingly compressible at sub-ion scales \cite{Kiyani2013,Chen2013}. Turbulence is also characterized by intermittency \cite{Matthaeus2015}. Observations and numerical simulations have shown that plasma heating is often located near intermittent coherent structures \cite{Wu2013}. Coherent structures exhibit multi-scale phase coherence \cite[e.g.][]{Koga2003,Hada2003,Perrone2016,Perrone2017,Roberts2016}. This is different to a plane wave which has phase coherence in time over the wave-packet's duration and is a single scale phenomena \cite[e.g.][]{Lion2016,Roberts2017b}. The level of intermittency is known to increase as the turbulent cascade progresses to smaller scales \cite{Sorriso-Valvo1999}, however, the nature of intermittency in the sub-ion range is unclear. In the Earth's magnetosheath \cite{Chhiber2018} and the solar wind \cite{Alexandrova2008c} observations show that the increase in the scale-dependent kurtosis persists into the sub-ion range. However, other observations in the solar wind \cite{Kiyani2009a} show this region is characterized by a non-intermittent monofractal scaling, and the scale-dependent kurtosis of fluctuations plateaus or even decreases \cite{Chhiber2018}. The reason is unclear but it has been suggested that high-frequency waves may have a randomizing effect on the increments \cite{Wan2012}, similar to how Alfv\'en waves at fluid scales can cause the transverse components to be less intermittent  \cite{Bruno2003}.
 
The nature of the fluctuations i.e. wave-like or structure-like has been a topic of spirited debate in the plasma turbulence community. The exact relationships between the two different phenomena and their interactions with one another remain unclear.  It is possible that both types of structures coexist \cite[e.g.][]{Roberts2013}, or that waves can generate structures \cite{Howes2016} increasing intermittency, or reduce it \cite{Bruno2003,Mallet2019}. Furthermore, structures can also generate waves themselves \cite{Karimabadi2013}. It is also often hypothesized that even in the highly nonlinear regime the linear physics remain important for the turbulence dynamics \cite{Goldreich1995,Mallet2015a,Groselj2019}.

The goal of this paper is to test the hypothesis that waves destroy the intermittency in the sub-ion range. The problem is tackled through a combination of wave-based analysis methods to identify and classify wave-like fluctuations and increment based methods that measure the intermittency. Both of these goals require data with high time resolution and sensitivity. Furthermore, to determine the wave-vectors multiple spacecraft are required, with distances of a few kilometers between them. This makes the data from the Magnetospheric MultiScale Mission (MMS) the only data set suitable to explore the problems described.

\begin{figure*}[htp]
    \centering
    \includegraphics[width=0.45\textwidth]{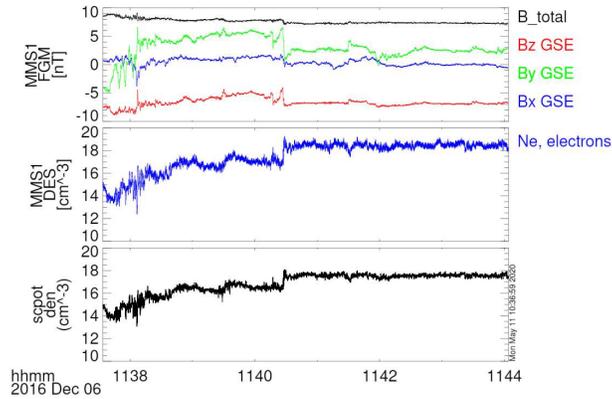}
    \caption{Measured time series from MMS from top to bottom. The magnetic field fluctuations from FGM, the electron density from FPI-DES. The spacecraft potential derived density from SDP.}
    \label{fig:fig1}
\end{figure*}

\emph{Data/Methods:} On December 6th 2016 at 11:37:34UT the MMS \cite{Burch2016} spacecraft record an interval of slow solar wind in burst mode until 11:44:04UT. During this time the four spacecraft are in a regular tetrahedral formation ensuring homogeneous spatial coverage with inter-spacecraft distances $\sim6$km. The mean values are as follows; the proton bulk speed is $350$km s$^{-1}$, the electron density is $16.9$ cm$^{-3}$, the magnetic field strength is 7.6 nT, the ion and electron plasma $\beta$ (ratio of thermal to magnetic pressure) values are 1.7 and 1.1 respectively. It has been reported that the Fast Plasma Investigation's Dual Ion Spectrometer (FPI-DIS) \cite{Pollock2016} may not have reliable measurements of the ion temperatures in the solar wind \cite{Bandyopadhyay2018}. The value of the ion $\beta$ using the mean temperature from OMNI data \cite{King2005} is 0.5. Figure \ref{fig:fig1} shows the measured data from the MMS, with the magnetic field being measured by the fluxgate magnetometer (FGM) \cite{Russell2016} at a sampling rate of 128Hz in the Geocentric Solar Ecliptic (GSE) coordinate system. In this coordinate system x points from the Earth towards the Sun and z points to the ecliptic north. The electron density is measured using two different methods, the first being a direct measurement from the FPI's Dual Electron Spectrometers (FPI-DES) \cite{Pollock2016} which have a sampling rate of 33Hz. The other method is to use the spacecraft potential which is measured with a sampling rate 8.192kHz from the spin plane double probes (SDP) \cite{Lindqvist2016}. We calibrate the spacecraft potential measurement using lower resolution electron density from FPI to give a higher time resolution measurement of the electron density \cite{Pedersen1995,Roberts2017}. The procedure to infer density from the spececraft potential has been used by several authors using MMS data \cite[e.g.][]{Andriopoulou2018,Graham2018,Torkar2019,Roberts2020a}. A detailed description of the application of this method in the solar wind with MMS is available in \cite{Roberts2020}.

 The spacecraft potential is governed by the currents to and from the spacecraft. These include, the electron thermal current to the spacecraft, the photoelectron current from the spacecraft and several other currents. In a cold, sparse plasma such as the solar wind the contributions of other currents are typically negligible and the photoelectron current and the electron thermal current are approximately equal and oppositely directed. The electron thermal current can be estimated from the plasma measurements, 
 
 \begin{equation}
    I_{e}=-A_{\text{spac}}qn_{e}\sqrt{\frac{k_{B} T_{e}}{2 m_{e} \pi}} \left(1+\frac{q  V_{sc}}{k_{B}T_{e}}\right)
    \label{Eq2}
\end{equation}

Where $A_{spac}$ is the spacecraft area, $V_{sc}$ is the spacecraft potential, $q$ is the fundamental charge, $m_{e}$ is the electron mass, $T_{e}=11.0$eV is the measured electron temperature from FPI-DES for this interval and $k_{B}$ is the Boltzmann constant. The photoelectron current is modeled as an exponential function;
 
 \begin{equation}
     I_{\text{ph}}=I_{\text{ph0}}\exp{(-V_{sc}/V_{0})},
 \label{photoelectron}
 \end{equation}
 
 where $I_{\text{ph0}}$ and $V_{0}$ are the photoelectron currents and potentials obtained by fitting the electron thermal current from FPI-DES to the spacecraft potential \cite[e.g][]{Pedersen1995}. A table of the constants used for calibrating this interval are given in Tab \ref{photo}.
 
 \begin{table}[h]
 \caption{Table showing the photoelectron parameters used for converting the spacecraft potential into a density}
     \centering
     \begin{tabular}{lcc}
          & $I_{ph0}$ ($\mu$ A)&$V_{0}$ (V) \\
          MMS1&301.5&2.0\\
          MMS2&474.2&1.7\\
          MMS3&375.9&1.9\\
          MMS4&219.9&2.1\\
     \end{tabular}
     
     \label{photo}
 \end{table}
 
 By equating Eq \ref{photoelectron} and \ref{Eq2} and solving for the electron density we obtain;

    \begin{equation}
    n_{e,SC}=\frac{1}{qA_{\text{spac}}} \sqrt{\left(\frac{2\pi m_{e}}{k_{B}T_{e}}\right)}\left(1+\frac{qV_{sc}}{k_{B}T_{e}}\right)^{-1}I_{ph}
    \label{neest}
\end{equation}

The calibration procedure and the methodology for removing the spin is discussed in more detail in \cite{Roberts2017,Roberts2020}. A comparison of the electron density derived from the spacecraft potential and the direct measurement from the FPI-DES is shown in Fig \ref{fig:fig1}. The effects of charging timescales can be neglected as we consider frequencies much lower than the charging timescales which are of order a few kilo Hertz \cite[e.g.][]{Chen2012b} in the solar wind. For an estimation of the timescale we assume the photoelectron flux is constant and therefore any changes in the spacecraft potential will be due to the changes in the ambient plasma. The photoelectron flux is dependent on the UV flux from the Sun, primarily the Lyman Alpha emission \cite[e.g.][]{Lybekk2012}. Which we expect to be constant during the short interval studied here. The photoelectron emission can vary due to strong electric fields ($\gtrsim 40 $mV/m) \cite{Torkar2017,Graham2018,Roberts2020a}, however, such large electric fields are associated with magnetopause crossings, not quiet solar wind. Therefore this assumption is justified. Following the calculation of \citet{Chen2012b} the charging timescale in the solar wind can be approximated by $\delta t=(CV_{0})/I_{e}$, where $C$ is the total capacitance of the spacecraft, which is estimated to be 2nF from the sum of the capacitances of each component of the spacecraft \cite{Graham2018}, and $I_{e}$ is the electron photocurrent which in the solar wind is typically of the order of $30-40\mu$ A. This corresponds to a frequency of $\sim7$kHz. An estimate of the charging time scale of the MMS spacecraft in the solar wind using a dust impact found that it was of the order of micro seconds \cite{Lhotka2020}. Both of these estimates correspond to frequencies that are much larger than the Nyquist frequency of the Burst mode data and much larger than the instrumental noise which occurs near 40Hz \cite{Roberts2020}. Therefore, the charging timescale will not affect our results.

%
%
%

To obtain spatial information about a plasma from single point measurements Taylor's hypothesis is often applied. This assumes the fluctuations evolve on a timescale longer than advection time over the measurement point so that the spacecraft sees a one-dimensional cut through the plasma. Should fluctuations be dispersive this hypothesis breaks down, so that multi-point data analysis techniques become necessary. To measure the wave-vector anisotropy and the plasma frame frequencies of the fluctuations we apply the Multi-Point Signal Resonator technique (MSR) \cite{Narita2011}. This method uses data from four spacecraft and filters the signal for a given spacecraft frame frequency and wave-vector using the phase delay between spacecraft. This method assumes weak stationarity of the time series and that it can be mathematically described as a superposition of plane waves with random phases and a small component of incoherent noise. However, crucially, Taylor's hypothesis is not required.  

The MSR technique estimates the wave power in four-dimensional space $P(\omega_{sc}, \mathbf{k})$ for a spacecraft frame frequency $\omega_{sc}$ and a wave-vector $\mathbf{k}$. The peak in $P(\omega_{sc}, \mathbf{k})$ can be used to find the wave-vector associated with the highest energy density at a given spacecraft frame frequency. This wave-vector is Doppler shifted to the plasma frame via $\omega_{pla}=\omega_{sc}-\mathbf{k}\cdot\mathbf{v_{sw}}$ where $\mathbf{v_{sw}}$ denotes the mean ion bulk velocity. The $\omega_{pla}$-$k$ relation and the angle between $\mathbf{k}$ and the mean magnetic field direction can then be compared to linear solutions of the Maxwell-Vlasov equation. 

The accessible range of scales is determined by the inter-spacecraft separations $d_{sc}$, where a maximum wave-number is given as $k_{\text{max}}=\pi/\langle d_{sc}\rangle$ ; the Nyquist wave-number. This corresponds to an upper limit of $f_{\text{max}}=k_{\text{max}} (v_{sw}-v_{ph})/{2\pi}=(v_{sw}-v_{ph})/2\langle d_{sc}\rangle\sim 24$Hz, where $v_{ph}=\textbf{Max}(c_{s},v_{A})$. For this interval the sound speed $c_{s}=58$ km s$^{-1}$ exceeds the Alfv\'en speed $v_{A}=38$ km s$^{-1}$. The minimum wave-number and frequency is selected based on resolving the wave-vector with an accuracy of $10\%$ which we set to $k_{\text{min}}=k_{\text{max}}/25$ \cite{Sahraoui2010} for a plane wave. This corresponds to a minimum frequency of $f_{\text{min}}\sim 1$Hz. We use the limit of $f_{\text{min}}\sim 2$Hz as some of the recovered associated wavenumbers with frequencies below 2Hz have large relative errors. We note that the magnetic field measurement on MMS is not sufficiently sensitive to be used for the majority of the frequency range in the solar wind and becomes noisy near $f_{sc} \gtrsim 5$Hz, whereas the density data become noisy at $f_{sc} \gtrsim 40$Hz and have the necessary time resolution and sensitivity to cover the range [$f_{\text{min}},f_{\text{max}}$]. Before inputting the data into the MSR method the density data from all spacecraft are resampled onto the same timeline, at a reduced rate of 5378.31Hz giving $2^{21}$ data points which are split into 128 windows of 16384 data points which overlap by $50\%$ giving a total of 256 windows. There are some instrumental effects in the spacecraft potential data at 13-14 Hz which are discarded from our study.

%
%
%
\begin{figure*}[htp]
    \centering
    \includegraphics[width=0.85\textwidth]{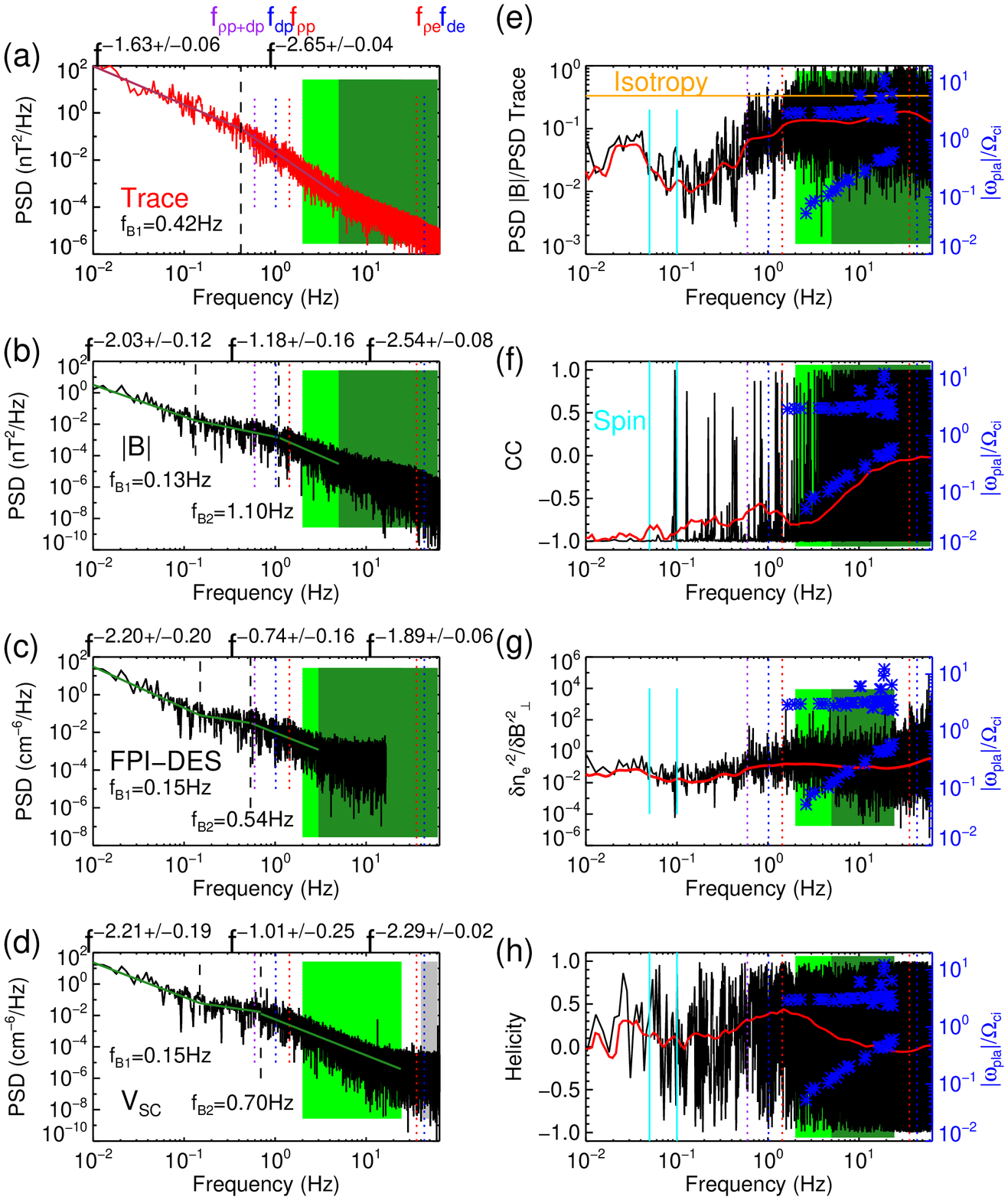}
    \caption{(a) Fourier Power spectral densities (PSD) of the trace magnetic fluctuations (b) PSD of the electron density fluctuations from FPI-DES. (c) PSD of the magnetic field magnitude fluctuations (d) PSD of the electron density fluctuations from the spacecraft potential (e) The ratio of the PSDs of magnetic field magnitude fluctuations to the trace fluctuations. (f) The cross-correlation spectra of $B$ and $n_{e}$. (g) the ratio of normalized electron density and perpendicular magnetic fluctuations (h) the magnetic helicity spectra. In figures (e-h) the black lines denote the quantities obtained from Fourier analysis while the thick red curves denote the estimation of the quantities based on wavelet analysis. The green and dark green areas denote [$f_{\text{min}},f_{\text{max}}$]. The dark green area denotes the region where FGM reaches the noise floor but where the density measurement remains viable. The second y-axis in (e-h) shows the plasma frame frequencies from the MSR technique which are overplotted for comparison with the spectral estimates.}
    \label{fig:fig2}
\end{figure*}

 \emph{Results Spectral Estimators:} Figure \ref{fig:fig2}(a-d) shows the Fourier power spectra of the magnetic and density fluctuations. The dotted lines denote the Taylor shifted gyroradius  $f_{\rho p,e}=v_{sw}/2\pi\rho_{p,e}$ and inertial length ,$f_{d p,e}=v_{sw}/2\pi d_{p,e}$ for protons and electrons and $f_{\rho _{p}+d_{p}}$ denotes the combined scale which for the protons is associated with cyclotron resonance \cite{Bruno2014}. Here $d_{p,e}$ is the inertial length where the particles decouple from the magnetic field and $\rho_{p,e}$ is the gyration radius of the particles about the magnetic field direction. The trace magnetic fluctuations show two power laws, while the compressive fluctuations from three different measurement methods show a flattening between the ion inertial and kinetic scales \cite{Chandran2009}. To estimate the compressive magnetic fluctuations we use fluctuations in the magnitude of the magnetic field \cite[e.g.][]{Perrone2016}. The density fluctuations are shown from both FPI-DES and the spacecraft potential method. The FPI measurement flattens at $f_{sc}>3$ Hz due to instrumental noise while the spacecraft potential measurement can be used throughout the entire range of interest before becoming noisy at 40Hz. The green shaded area denotes the scales of interest of this study [$f_{\text{min}},f_{\text{max}}$].The dark green area denotes frequencies which are in the scales of interest but where the FGM or FPI-DES signals are at the noise level. The noise level of the spacecraft potential is outside the scales of interest and denoted by the grey shaded area. There are no indications from the Fourier spectra that there are any spin tones or any higher harmonics present although in the cross correlation spectra in Figure \ref{fig:fig2}f there is a strong correlation near the second harmonic of the spin tone. Therefore there may be some unphysical effect on the cross correlation however these are very narrow in frequency and we expect the limitation to be smaller as the frequency increases. Typically, magnetic field measurements on MMS have spin tones at the first and second harmonics. The measured spacecraft potential in the solar wind usually shows spin tones up to 1Hz due to the sunlit surface changing as the spacecraft spins. However these have been removed before calibrating to density. Some residual effect may exist at some higher harmonics of the spin tone however this is unlikely to affect our measurements as our frequency range of interest is above 2Hz which is at the 40$^{th}$ harmonic of the spin tone.
 
 The ratio of the PSD of the magnetic field magnitude fluctuations to the trace fluctuations is shown in Figure \ref{fig:fig2}e. This gives a measure of the magnetic compressibility which increases in the sub-ion range approaching a value of 1/3. A value of 1/3 indicates equal power in the compressible and the two transverse components and is denoted by the orange line. The black curves denote the estimation from Fourier analysis and the thick red lines denote the estimation from wavelet analysis \cite{Torrence1998} which has been averaged in time. In figures \ref{fig:fig2}e-g a second y-axis is presented showing the spacecraft frame frequencies obtained from the MSR analysis which is discussed later. These points show two distinct populations of waves, and are presented here for comparison with the spectral estimators.
 
 Figure \ref{fig:fig2}f shows the cross-correlation spectra;
 \begin{equation}
 CC=\frac{\mathfrak{R}(\tilde{n}_{e} \tilde{B}^{*})}{|\tilde{n}_{e}||\tilde{B}|}
 \label{CCEqn}
 \end{equation}
 
of the density and the magnetic field magnitude where the tilde symbols denote the Fourier transform (or the wavelet transform) and the asterisk denotes the complex conjugate. Using this definition, -1 denotes anti-correlation and +1 denotes correlation. At inertial scales, the fluctuations are anti-correlated, consistent with slow waves or pressure balanced structures \cite{Verscharen2017}, at smaller scales the anti-correlation persists consistent with the polarization of the kinetic Alfv\'en wave. There are also narrowband regions in frequency which exhibit positive correlations. In the low-frequency range, these appear at the second harmonic of the spacecraft spin (the spacecraft spin frequency and the second harmonic are indicated by the vertical cyan lines) and are unlikely to be physical. At larger frequencies there are some regions with positive correlations, these cannot be resolved by the wavelet spectrum as the frequency resolution is too low. A positive correlation is a characteristic of the magnetosonic family of waves rather than the Alfv\'en wave family. Some higher frequency points in blue may be associated with positive correlations but it is difficult to make this distinction especially as the majority of the high frequency points are in the region where the magnetic field is noisy $f_{sc}>5$Hz.  Figure \ref{fig:fig2}g shows the ratio of normalized density to perpendicular magnetic fluctuations \cite{Chen2013,Roberts2018b}. The kinetic Alfv\'en normalization is as follows;
 
 \begin{equation}
     \delta \mathbf{B}'=\delta \mathbf{B}/B_{0},
 \end{equation}
 
 \begin{equation}
     \delta n_{e}'=\left(1+\frac{T_{i}}{T_{e}}\right)^{1 / 2} \frac{c_{s}}{v_{A}}\left[1+\left(\frac{c_{s}}{v_{A}}\right)^{2}\left(1+\frac{T_{i}}{T_{e}}\right)\right]^{1 / 2} \frac{\delta n_{e}}{n_{0}},
 \end{equation}

This allows for the transverse components of the magnetic field to be estimated from
 \begin{equation}
 \delta \mathbf{B}_{\perp}^{'^2}=\delta \mathbf{B}^{'^2}-\frac{\left(c_{s}^{2} / v_{A}^{2}\right)\left(1+T_{i} / T_{e}\right)}{1+\left(c_{s}^{2} / v_{A}^{2}\right)\left(1+T_{i} / T_{e}\right)} \delta n_{e}'^{2}.
 \end{equation}
 
This relation comes from the nonlinear equations for kinetic Alfv\'en waves \cite{Schekochihin2009,Chen2013}. Here we use the ion temperature measurement from OMNI and the other parameters from MMS. For kinetic Alfv\'en waves the amplitudes of fluctuations normalized in this manner are expected to be approximately equal $\delta B_{\perp}'\simeq\delta n_{e}'$. For Fast/Ion Bernstein waves this $\delta B_{\perp}'\gg\delta n_{e}'$ is expected. The mean value of $\delta B_{\perp}'^2\gg\delta n_{e}'^2$in the scale of interest (before instrumental noise becomes significant) is 0.13 consistent with the results of \cite{Chen2013}, although there are regions where the individual Fourier modes differ substantially from the mean. Magnetosonic fluctuations would be expected to be smaller by more than one order of magnitude.
 
Figure \ref{fig:fig2}h shows the magnetic helicity which is defined as;
 
 \begin{equation}
     \sigma_{m}=\frac{2\mathfrak{I}\left(\tilde{B_{y}}\tilde{B_{z}}^{*}\right)}{|\tilde{B_{y}}|^{2}+|\tilde{B_{z}}|^{2}},
     \label{helicityeqn}
 \end{equation}

where the $y$ and $z$ subscripts denote the GSE components. This method assumes the wave-vector points into the $-x$ GSE direction. This assumption will be justified later with our MSR analysis. The helicity is a measure of the spatial rotation sense of the magnetic fluctuations about the wave-vector direction. This can be used as a diagnostic for the type of plasma wave. The helicity shows an increase before our region of interest which is commonly interpreted as being due to kinetic Alfv\'en waves when the angle $\theta_{BV}\sim90^{\circ}$ \citep[e.g.][]{He2011,Podesta2011,Roberts2015,Woodham2018}. However, at frequencies larger than 5Hz, the helicity becomes zero. This is likely due to instrumental noise. Low helicity could be is a characteristic of Ion Bernstein waves \citep{He2012}. Alternatively, the contributions of oppositely directed KAWs can create a mean helicity signal of zero if their power is balanced. We also note that fluctuation levels in the slow solar wind are naturally weaker and may thus create a weaker signal \cite{Bruno2015}. Nevertheless, our measurement of the helicity suggests that fluctuations are KAW-like in the range $f_{sc} \in [0.5-5]$Hz. At higher frequencies due to noise the measurement is inconclusive.
 
\begin{figure}
     \centering
     \includegraphics[width=0.45\textwidth]{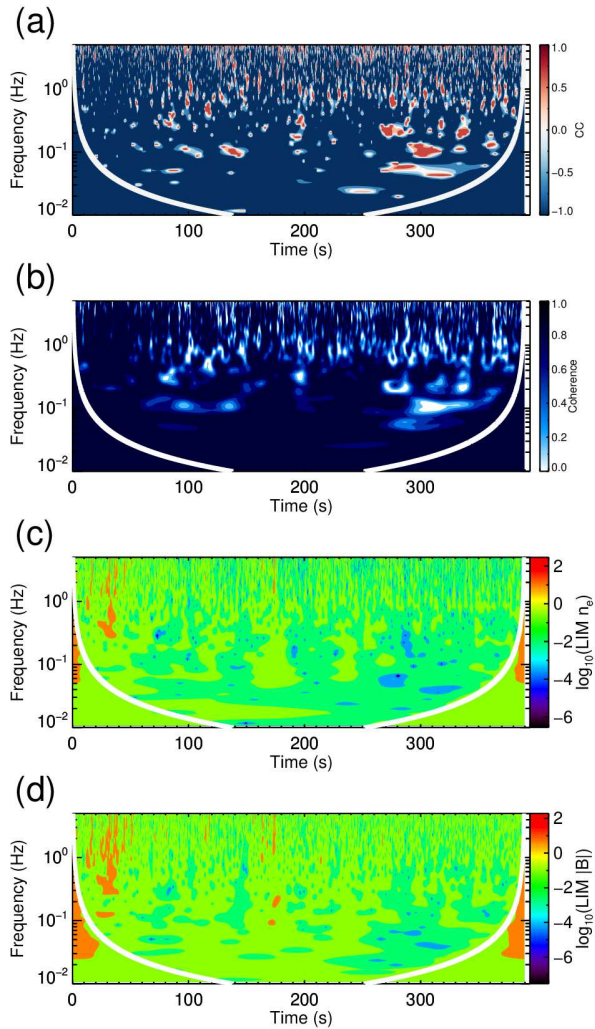}
     \caption{(a) wavelet cross correlation spectra between $|B|$ and $n_{e}$ (b) wavelet coherence spectra between $|B|$ and $n_{e}$ (c) Local intermittency measure for $n_{e}$ (d) local intermittency measure for $|B|$. The white line denotes the cone of influence, where below this line the results are unreliable due to edge effects.}
     \label{fig:crosscorrelation}
 \end{figure}

Figure \ref{fig:crosscorrelation} shows the wavelet spectra of the cross-correlation, the coherence, and the local intermittency measure for both electron density and magnetic field magnitude. The local intermittency measure (LIM) is defined as;
 
 \begin{equation}
    LIM(t,\tau)=\frac{|\tilde{n}_{e} (t,\tau)|^2}{\langle|\tilde{n}_{e}(t,\tau)|^2\rangle_{t}}
\end{equation}
 
where the tilde denotes the wavelet transform, which is presented for both $n_{e}$ and $|B|$ in figures \ref{fig:crosscorrelation} c and d respectively. 
 In Fig \ref{fig:crosscorrelation}a we see a similar picture to the global measurements presented in Figure \ref{fig:fig2}f with anti-correlations dominating. However, there are small regions where positive correlations are present. Rather than a different wave mode being present in these regions, the wavelet coherence in \ref{fig:crosscorrelation}b and the local intermittency measure suggest that these are regions of low amplitude fluctuations where the coherence is weak. The strong anti-correlation, the amplitudes and the magnetic helicity suggest that KAW like fluctuations dominate in this range. However, spectral estimators that use the magnetic field are limited to 5Hz due to noise and we cannot infer conclusions from these estimators at higher frequencies. Instead, we now use the electron density measurements which have sufficient time resolution and sensitivity to study at frequencies far into the sub-ion range $f_{sc}>$5Hz.

\begin{figure*}[ht]
    \centering
    \includegraphics[angle=-90,width=0.95\textwidth]{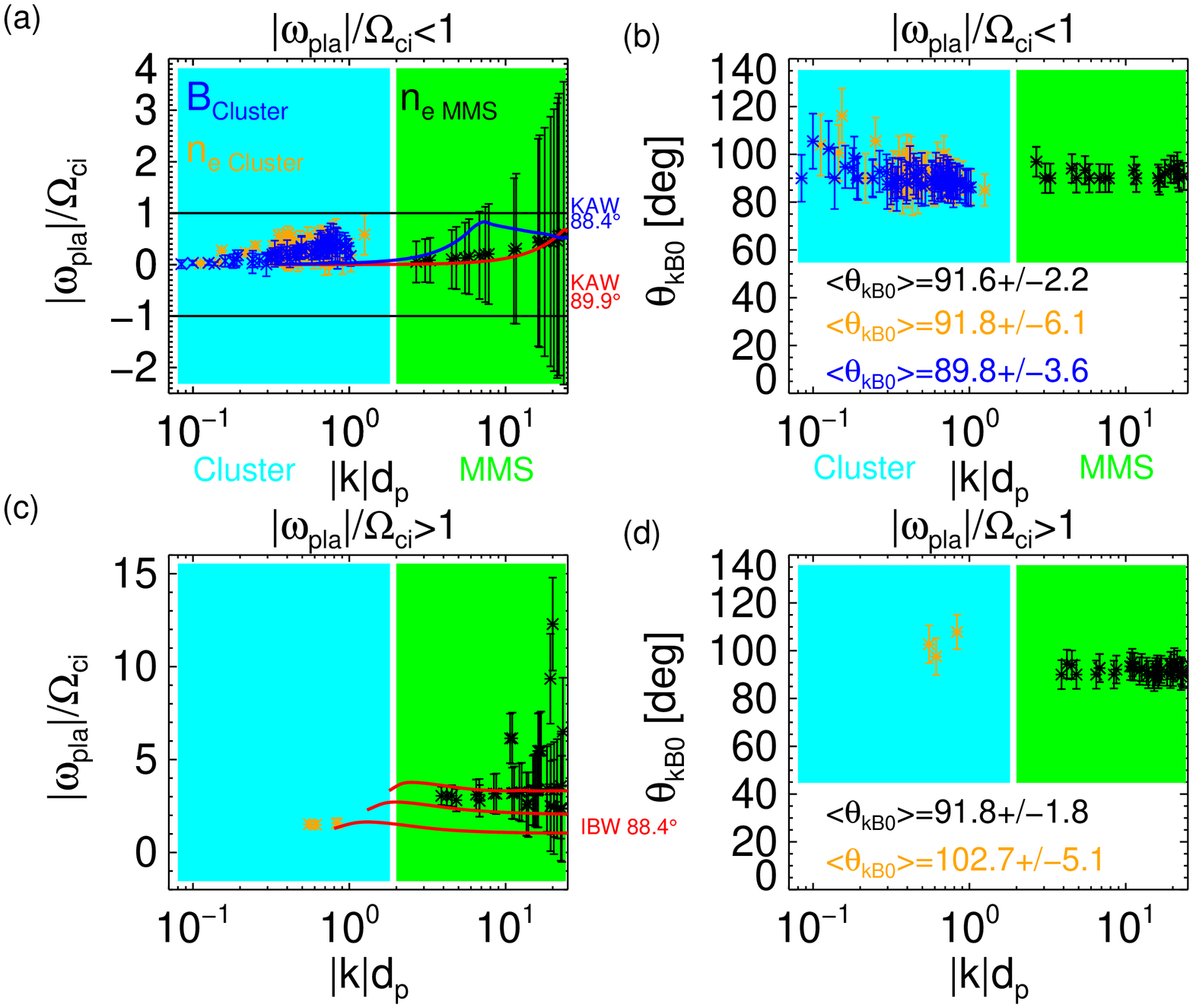}
    \caption{Dispersion relations and propagation angles obtained from the MSR technique for plasma frame frequencies $<\Omega_{ci}$ (a,b) and $>\Omega_{ci}$ (c,d). The blue (orange) points denote the magnetic field (density) data points from Cluster for solar wind with similar conditions. The black points denote the data from our MMS interval. The errors are based on the uncertainty in the wave-vector and the standard deviation of the ion velocity. The lines show the solutions to the hot-plasma dispersion relations. }
    \label{fig:fig3}
\end{figure*}

 \emph{Results: MSR analysis} Figure \ref{fig:fig3}a,c shows the dispersion relation diagrams and Fig \ref{fig:fig3}b,d shows the angle the wave-vectors make with the mean magnetic field direction calculated over the interval for the density measurements.  For clarity, these have been split into data points which are below the proton cyclotron frequency $\Omega_{ci}$ and those above. The results show a mixture of plasma waves both above and below the ion cyclotron frequency consistent with previous studies using the magnetic field \citep{Narita2011,Narita2016,Gershman2018}. 
 
 For the context of the larger scales, we plot data points reported in \cite{Roberts2017}. The study of \cite{Roberts2017} used Cluster \citep{Escoubet2001} magnetic field and spacecraft potential data with inter-spacecraft distances of 200km and the plasma parameters are comparable to the MMS interval. The wave-vectors from both intervals are shown to make an almost perpendicular angle with the magnetic field, consistent with other studies at larger scales \cite{Sahraoui2010a,Narita2011a,Roberts2013,Roberts2015a,Narita2016,Gershman2018}.

%
%
%

\begin{figure}[ht]
 \includegraphics[width=0.45\textwidth]{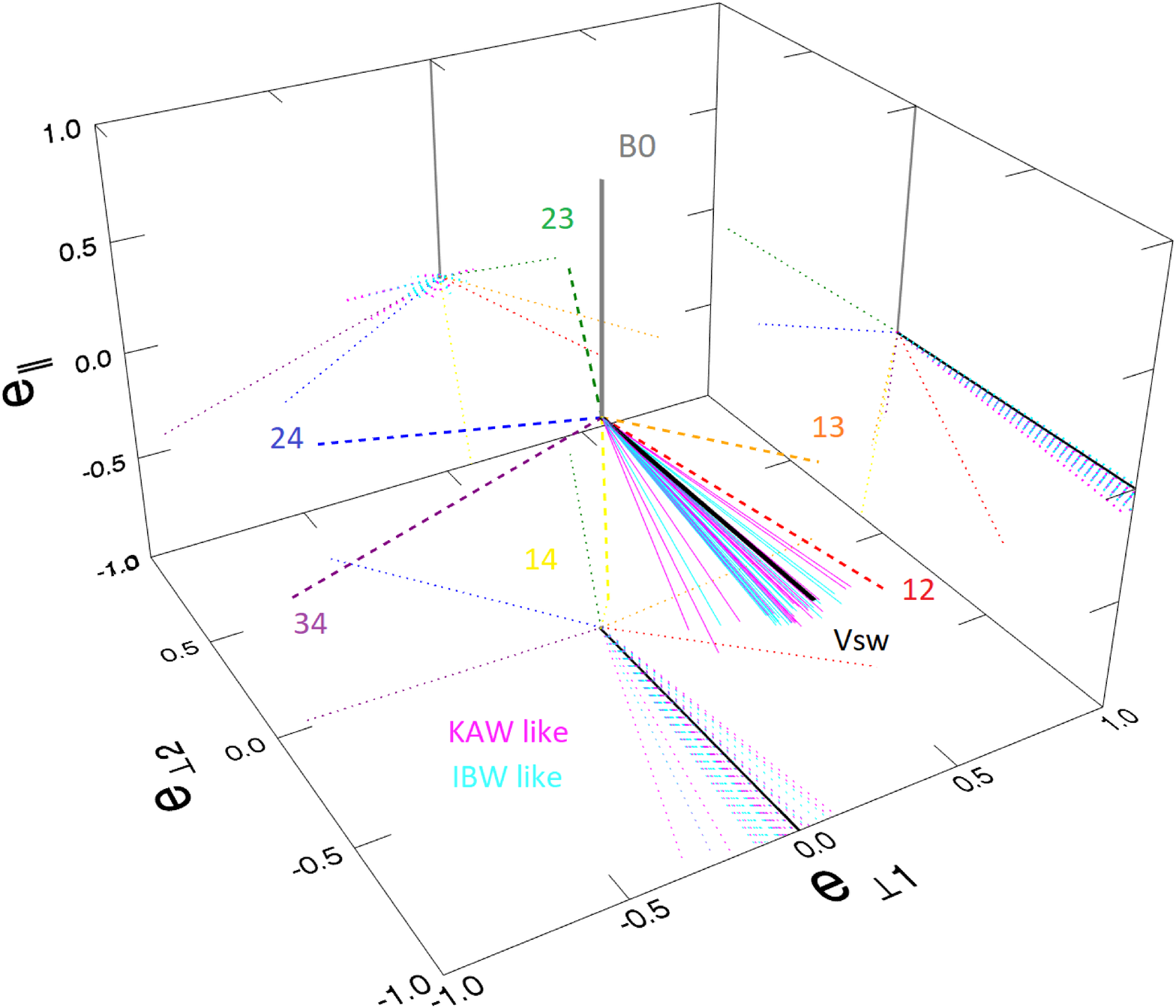}
 \caption{Three dimensional representation of the different directions relevant for this study. The grey line denotes the mean magnetic field direction in the $\mathbf{e}_{\parallel}$ direction. The thick black line denotes the mean bulk velocity direction which is approximately perpendicular to the magnetic field direction and is predominantly in the negative $\mathbf{e}_{\perp 2}$ direction. The thick dashed lines denote the unit vectors of different inter-spacecraft baselines. The thin cyan and magenta lines denote wave vectors of the recovered IBWs and KAWs respectively which make an approximately perpendicular angle with the mean magnetic field direction and a small angles $\sim 10^{\circ}$ with the bulk flow direction. For better presentation projections of the different unit vectors are shown on the different planes.}
 \label{fig:geometry}
 \end{figure}

 The Cluster data show that the fluctuations have low frequencies compared to $\Omega_{ci}$ consistent with KAW turbulence, and the slightly larger spread in the density points was interpreted as being due to a superposition of KAWs and kinetic slow waves \cite{Roberts2017,Safrankova2019}. We note that advected structures with no intrinsic plasma-frame frequencies would exhibit a similar low-frequency signature. Structures such as magnetic holes or pressure balanced structures would also be able to explain the anti-correlations between density and magnetic field magnitude. The lower frequency fluctuations from MMS presented in Fig \ref{fig:fig3}a show that there is an increase in the frequency at the smaller scales consistent with the KAWs. This is plotted along with the dispersion relation of KAWs obtained from the New Hampshire Dispersion relation Solver \cite{Verscharen2018} for the measured plasma conditions (where the OMNI ion $\beta$ is used) and the measured mean propagation angle of $88.4^{\circ}$. The dispersion relation for the KAW at an angle of $89.9^{\circ}$ is also plotted and shows better agreement with the data. We emphasize that this plotted angle of $89.9^\circ$ is within the standard deviation of the propagation angle. Kinetic Alfv\'en waves are very sensitive to the propagation direction and an uncertainty of a few degrees has a large impact on the solutions from linear theory. The wave-vector of the fluctuations also make a small angle with the ion bulk velocity direction of about $\sim 10^{\circ}$. Figure \ref{fig:geometry} shows the unit vectors of the recovered wave-vectors at each frequency and the unit vectors of the mean magnetic field direction, the mean bulk velocity, and the inter-spacecraft baselines. This figure is presented in mean field aligned coordinates with $\textbf{e}_{\parallel}=\mathbf{B}/B_{0}$, $\textbf{e}_{\perp 1}=\textbf{e}_{\parallel}\times\mathbf{V_{sw}}/|\mathbf{V_{sw}}|$. So that the velocity is primarily in the negative $\textbf{e}_{\perp 2}$ direction. This demonstrates that the wave-vectors associated with the most energetic fluctuations are almost perpendicular to the mean magnetic field and are in the direction of the mean bulk velocity. This observation justifies the assumption used for measuring the magnetic helicity for this interval.
 
\begin{figure}[ht]
    \centering
    \includegraphics[angle=-90,width=0.48\textwidth]{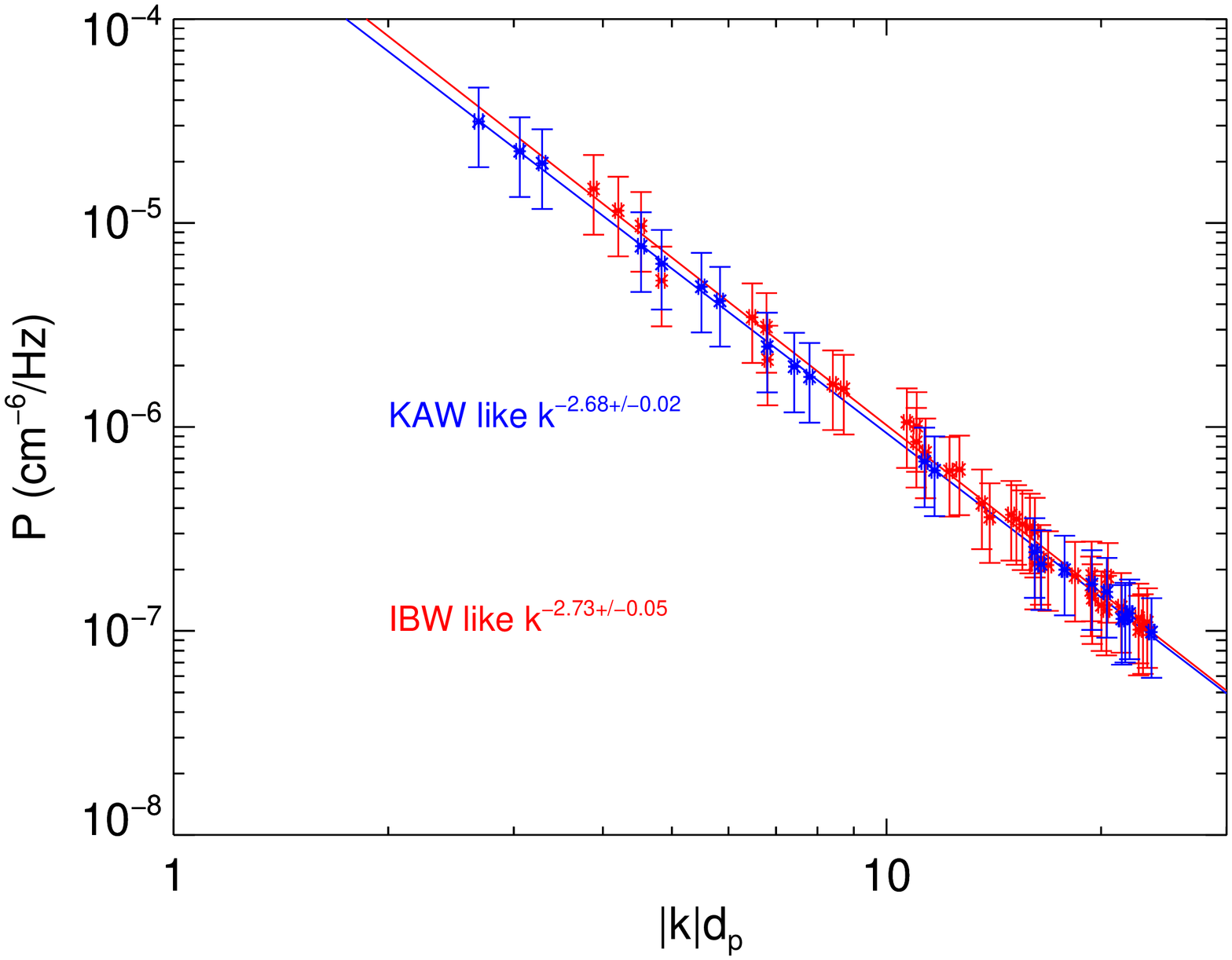}
    \caption{Peak power of the fluctuations at the recovered wave-vectors. Red denote the IBWs while blue denote KAWs. The vertical error bars are obtained from the $95\%$ confidence interval for 1024 degrees of freedom (256 windows, 4 spacecraft), the horizontal error bars are determined for a plane wave as per \cite{Roberts2017}. Power law fits are performed to both fluctuation types and the error is obtained from the residuals of the least squares fit.}
    \label{fig:fig7}
\end{figure}

 For frequencies larger than the cyclotron frequency the fluctuations occur predominantly near the third harmonic of the cyclotron frequency. This suggests that some fluctuations follow the dispersion relation of IBWs \cite{Podesta2012}. In figure \ref{fig:fig7} the peak power of the fluctuations is obtained and is displayed as a function of the wavenumber. Both the KAW like fluctuations and the IBW like fluctuations have comparable power across all scales, however the IBWs have a slightly steeper spectral index. 
 
\begin{figure}[ht]
    \centering
    \includegraphics[width=0.48\textwidth]{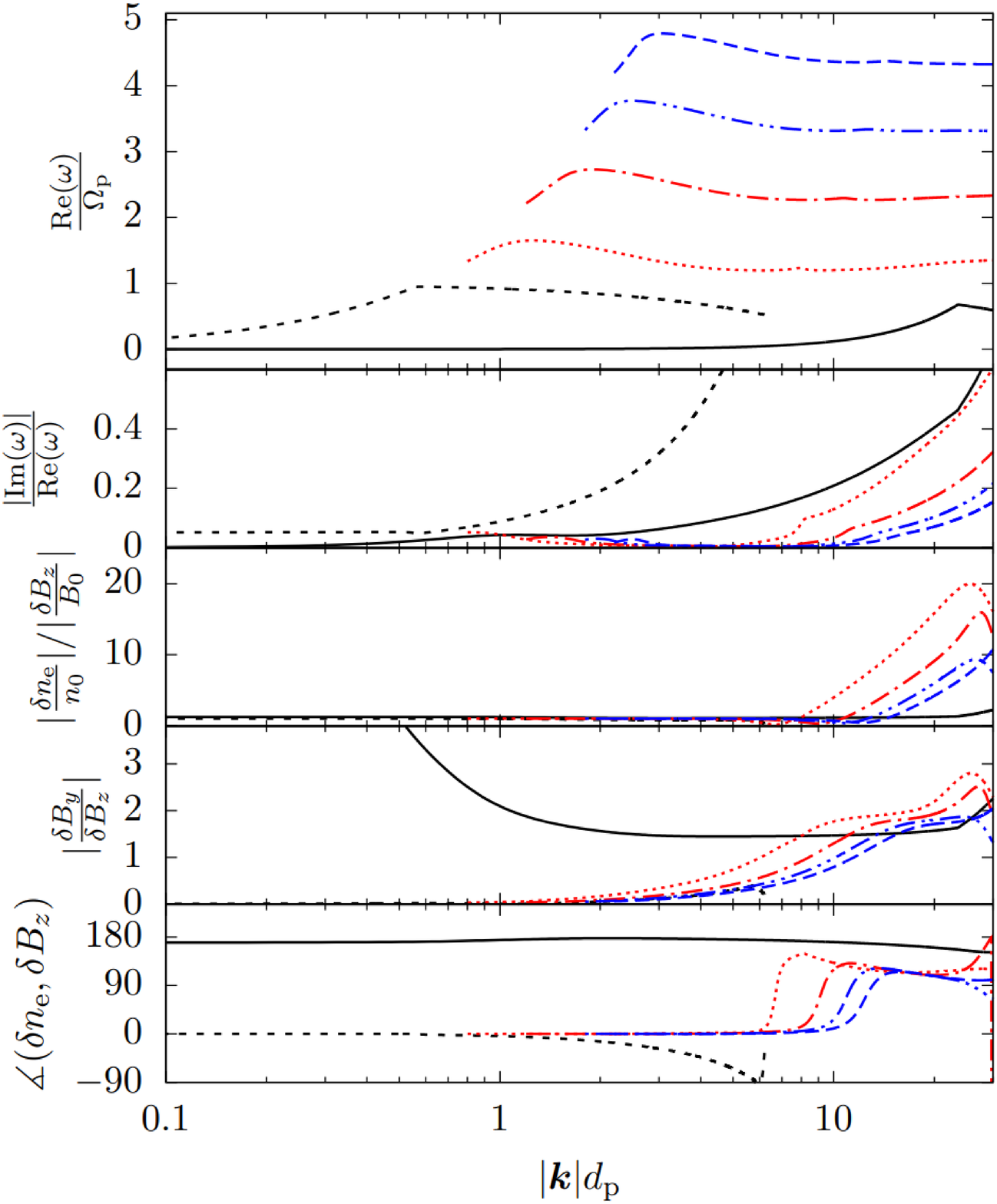}
    \caption{Linear Theory properties of the KAW (solid black line) and the magnetosonic/Ion Bernstein branches, from top to bottom. The dispersion relations (real part of the frequency), the damping rates, The ratio of density to compressible magnetic amplitudes, the magnetic compressibility, the phase angle between the density and compressible magnetic flcutuations.}
    \label{LinearTheory}
\end{figure}

Figure\ref{LinearTheory}, shows the properties of the linear kinetic Alfv\'en waves and several harmonics of the ion Bernstein waves obtained from NHDS \cite{Verscharen2018}. The dispersion relations, linear damping rates, compressibility, magnetic compressibility, and the phase angle between the magnetic field and density fluctuations is shown. Here $\pm180^{\circ}$ corresponds to anti-correlation, while $0^{\circ}$ corresponds to a positive correlation. The large damping rate of the IBW at the proton cyclotron frequency possibly explains why we do not measure any points at this frequency in Fig\ref{fig:fig3}. Their linear damping rate is too large to exist for more than a few wave periods. This could mean that the damping rates are much stronger than any driving process so that they cannot be generated, or that they do not exist as they are not excited at all. However, higher-order ion-Bernstein branches are less severely damped and thus able to exist. For these plasma conditions i.e. highly oblique wave-vectors, and the plasma $\beta$ values the waves tend to avoid crossing $\omega=\Omega_{ci}$ as the damping rates become large for these waves. This is different from the solutions presented in \cite{Sahraoui2012} where the solutions cross $\omega=\Omega_{ci}$ for smaller propagation angles ($\theta_{kB}<88^{\circ}$), for a larger plasma $\beta$ and a larger ion to electron temperature ratio than is measured here. The compressibility of both the KAWs and the IBWs with $k_{\perp}\gg k_{\parallel}$ increase at larger wavenumbers which are consistent with the isotropy observed in the power of the compressive and transverse components observed in Figure \ref{fig:fig2}e, and in \cite{Kiyani2013}. For KAWs the density and magnetic field fluctuations are anti-correlated where the opposite is true for the ion Bernstein waves. However, in the sub ion range the relative power of density fluctuations compared to the compressive magnetic field is very large. Considering this observation linked with the fact that the magnetic field instrument is nearly at the noise floor it is difficult to identify positively correlated fluctuations through our spectral estimators.

%
%
%

\begin{figure}[ht]
    \centering
    \includegraphics[angle=-90,width=0.48\textwidth]{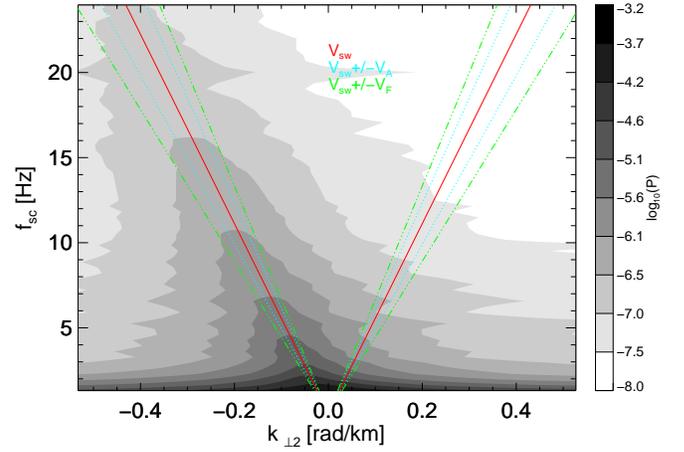}
    \caption{Power distribution of fluctuations in the  ($\perp 2$) flow direction and spacecraft frame and the flow direction. Also indicated on the plot are the bulk speed (red solid line) , the sum (and difference) with the Alfv\'en speed (cyan dotted) and the magnetosonic speed (green dot dashed).}
    \label{Dispersion2}
\end{figure}

The location of the wavenumber with the largest power poses an important limitation for our method. So this method preferentially selects those wavemodes with the largest power in the signal. In figure \ref{Dispersion2} we present a one-dimensional cut of the power spectrum through $k_{\perp 2}$, which is approximately the flow direction, at the smallest wavenumbers for the other two components. We find these one-dimensional cuts for different spacecraft frame frequencies and stack these results into the contour plot. We over-plot are the relation expected for an advected structure (based on Taylor's hypothesis), for an Alfv\'en wave ($v_{sw}+v_{A}$) and a fast magnetosonic wave ($v_{sw}+v_{F}$ where $v_{F}=\sqrt{v_{A}^2+c_{s}^2}$). Although the largest power is in fluctuations with low phase speeds i.e. KAW and advected structures, there is a non-negligible component of power in between the curves for the fast magnetosonic speed. This suggests that while the low frequency components are dominant there are contributions from other types of fluctuations. These may be Ion Bernstein modes as suggested in Fig \ref{Dispersion2}, however there could be contributions from sideband modes \cite[e.g.][]{Perschke2014,Narita2018}, where wave-wave interactions have broadened the power distribution. These side-band fluctuations may retain the properties of their parent waves e.g. polarizations, fluctuation amplitudes however their frequencies are different to their parent waves frequencies. Another possibility is that the broadening is due to fluctuations in the velocity. In the Doppler shift equation large fluctuations in the velocity can cause broadening in frequency \cite[e.g.][]{Narita2014a,Perschke2016}.

\begin{figure}
    \centering
    \includegraphics[angle=-90,width=0.45\textwidth]{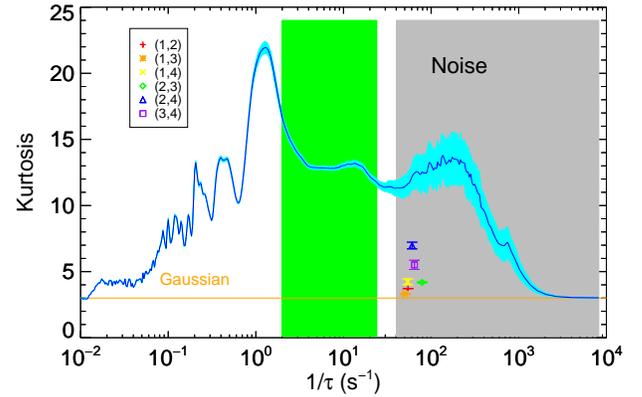}
    \caption{Scale-dependent kurtosis of the density fluctuations. The cyan shaded area denotes $3\sigma$ calculated from one hundred bootstrap re-samplings. The green area denotes the timescales of interest. The points denote the kurtosis calculated from spatially lagged spacecraft pairs. Similarly the error bars denote the $3\sigma$ error.}
    \label{fig:fig4}
\end{figure}

 \emph{Results:Intermittency} Figure \ref{fig:fig4} shows the scale-dependent kurtosis which is defined for increments of the measured density as;
 
 \begin{equation}
 \kappa(\tau)=\frac{\langle \delta n_{e}^4 \rangle}{\langle \delta n_{e}^2 \rangle^2}.
 \label{kurtEQN}
 \end{equation}
 
  where the increments are calculated from time lags $\delta n_{e}(t,\tau)=n_{e}(t+\tau)-n_{e}(t)$ from a single spacecraft or a spatial lag between two spacecraft $\delta n_{e}(\lambda_{1,2})=n_{e}(\lambda_{1})-n_{e}(\lambda_{2})$. The single spacecraft measurement gives a measure of the fluctuations along the bulk flow direction, while different spacecraft pairs measure the fluctuations along different baselines.

  A kurtosis of 3 indicates non-intermittent Gaussian fluctuations and larger values indicate that there are heavier tails in the probability distribution function due to intermittent large amplitude fluctuations. We remove 4 seconds at the edges of the time interval to avoid any large fluctuations associated with the interpolation. We also remove outliers that are not sufficiently sampled in the signal that can have an undue influence on the value of the kurtosis due to the finite signal length. To do this we follow the scheme of \cite{Kiyani2006}, removing the largest fluctuation until the value of the kurtosis converges.
 
 The time-lagged kurtosis is seen to increase, reaching a maximum near proton scales but then decreases to a value $\sim 12$ in our range of interest and does not show a strong dependence on the timescale. At higher frequencies, instrumental noise and other effects such as dust strikes \cite{Escoubet2020} affect the measurement. Figure \ref{fig:fig4} also shows the kurtosis from the spatial lags indicated by the symbols. These are converted to a time lag by using the measured ion velocity. Although these values are in the noise region for the time lags, they are likely to be more robust to the effects of noise than the time lags \cite{Chhiber2018} as noise is uncorrelated between spacecraft. Generally the difference in the time and space lag could be due to a number of reasons. Its could represent a breakdown in Taylor's frozen in flow hypothesis due to the presence of waves or structures that have evolution time scales faster than the advection time over the spacecraft. This would be consistent with waves with a high phase speed such as Ion Bernstein waves which are part of the magnetosonic family of waves and could violate the hypothesis \cite[e.g][]{Klein2014a}. Alternatively, structures may merge or evolve on at faster timescales than the advection time. One possible instrumental limitation is that the instruments may have insufficient amplitude resolution to resolve the fluctuations between spacecraft. The root mean squared values of the density fluctuations for the different spacecraft are; $[(1,2),(1,3),(1,4),(2,3),(2,4),(3,4)]= [0.15,0.10,0.13,0.10,0.11,0.11]cm^{-3}$ while the RMS value of the time lag for the same scale from a time lag is $0.03cm^{-3}$. This is similar to what was observed in the magnetic field by \cite{Chhiber2018}. Therefore, as the fluctuations measured from the spatial lags are larger in terms of amplitude the measurement between the two spacecraft should be less susceptible to noise.  Another possibility is that the measured kurtosis is direction dependent \citep{Sorriso-Valvo2006,Sorriso-Valvo2010,Yordanova2015}. Lastly, a sampling effect related to the bulk flow direction may lead to a situation in which we sample more structures in this direction than along the spacecraft separation directions \cite{Turner2011,Lacombe2017,Matteini2020}.  We notice that different baselines have different values of kurtosis with the spacecraft pairs that are separated along the mean-field direction (1,2),(1,3),(1,4) having lower values than (2,3),(2,4), and (3,4). The geometry of this system is shown in fig \ref{fig:geometry}. This is consistent with either IBWs or KAWs randomizing the electron density fluctuations in the parallel direction, while in the direction of wave propagation there is little or no effect. The observations show that the kurtosis in the parallel direction is reduced and the small error bars give us confidence that this is reduction is physical. We note that a probe failure on MMS4 means that the spacecraft potential measurement is made with two probes rather than the four probes on the other spacecraft. Therefore there may be some effect that causes the pairs with spacecraft four to have different results. We do not expect the result to be affected here as the pair (2,3)  (predominantly in the flow direction) has a larger kurtosis than both (1,2) and (1,3) (parallel direction). 
 
%
%
%

 

  \emph{Summary/Conclusions;} we investigate compressive fluctuations in the sub-ion range of solar wind turbulence. At these scales, the density fluctuations exhibit strong wave-vector anisotropy, and plasma frame frequencies which are consistent with KAW fluctuations, but there is also a population of high-frequency fluctuations consistent with higher harmonics of the ion Bernstein mode. The power in both fluctuation types is comparable, however, the spectral index for IBWs is steeper than for KAWs. The measurements of the cross-correlation, magnetic helicity and the relative normalized amplitudes of density and magnetic fluctuations and the MSR analysis all support the interpretation that KAWs are present in this interval. The combination of all methods gives strong evidence that KAWs are an important component of solar wind turbulence. However, as the magnetic field becomes noisy near 5Hz we cannot confirm the presence of IBWs through the methods involving magnetic fields in the majority of the frequency range studied here. The presence of IBWs is only shown in our MSR analysis. This may be due to the instrumental limitations of the magnetic field measurement at high frequencies, or that wavelet analysis has insufficient frequency resolution to resolve the IBWs. This might also be due to the inherent nature of IBWs at small scales. The normalized density fluctuations are much larger than the compressive magnetic fluctuations, meaning that a cross correlation may not be able to reveal such fluctuations. Alternatively, this could be due to spatial aliasing \cite{Narita2009}, a limitation of wave-based analysis methods leading to the inability to differentiate between two wave-vectors where $\textbf{k}\cdot \textbf{r}$ that differ by an integer multiple of $2\pi$. Spatial aliasing can manifest itself in two ways; the first is that a wave near the edge of the analyzed k-space can cause a spurious peak which is also in the region of interest. Both the physical peak and the aliased peak would have the same power. We investigate each peak individually and find that this effect does not seem to be present. Practically we use the spacecraft configuration to define the region of k-space in which such a peak cannot cause an alias. This region of k-space is similar to the Brillouin zone in solid state physics \cite[e.g.][]{Neubauer1990,Motschmann1996,Glassmeier2001}. The other possibility for spatial aliasing is the existence of a wave outside of this zone that causes a peak inside; a possibility which we cannot completely rule out. However in turbulence we expect that waves at higher frequencies (and higher k) have less power and would not cause any significant aliasing. This hypothesis is supported by our results in Fig \ref{fig:fig2}. However we cannot completely rule out some aliasing effects especially at higher wavenumbers. One possibility is that the Ion Bernstein waves observed here are in fact KAWs that have undergone many wave-wave interactions. Wave-wave interactions can cause the frequency to broaden \cite[e.g.][]{Howes2013b,Narita2018}, which can mean that the frequency identification alone is a limited method of diagnosing wavemodes in a turbulent plasma. This would be consistent with the approximate equal power observed between the two modes in Figure \ref{fig:fig7}. In this scenario the IBW like fluctuations may in fact be KAWs whose frequency have been broadened by wave-wave interactions or by velocity fluctuations. This may explain why in the below 5Hz why we do not see strong signatures of IBWs. The low signal to noise ratio of the magnetic field measurement may result in a positively correlated signature however due to the low amplitude of one component may have a weak coherence (which is observed in Fig \ref{fig:crosscorrelation}ab). The MSR analysis of the density time series allows these fluctuations to be revealed. Indeed according the power in k-space as a function of k in the flow direction and the spacecraft frame frequency in Fig \ref{Dispersion2} there is significant power not only for the zero frequency modes but also at a range which includes the fast magnetosonic speed. The fact that there is significant power which is not accounted for when taking the peak power should be kept in mind when interpreting dispersion plots such as those in Fig \ref{fig:fig3} and in the literature \cite[see a review of the results in][]{Narita2018}.
  
  Ion Bernstein waves may provide an additional channel for turbulent heating \cite{Podesta2012}. If confirmed, this finding would have a major impact on the applicability of Taylor's hypothesis as they are strongly dispersive. The waves are observed at scales where there is a reduction in the scale-dependent kurtosis. The kurtosis is smaller in the direction along the mean magnetic field direction when compared to the flow direction. This is consistent with the interpretation that in the solar wind wave activity has a randomizing effect on the fluctuations leading to a reduction in the observed kurtosis. Instability driven waves can also destroy intermittency in this manner, however the KAWs discussed here are more likely generated by the cascade. We propose that at large scales Alfv\'en waves destroy the intermittency in the transverse components, as the cascade proceeds and waves develop large $k_{\perp}$ they become more compressive and in the sub-ion range begin to also affect the compressive intermittency. An outstanding question is how waves from the turbulent cascade can cause a reduction in intermittency. One possibility is that stochastic heating \cite{Chandran2010} changes the distribution of the fluctuation amplitudes, which is predicted to lead to a decrease of the scale-dependent kurtosis at around the dissipation scale \cite{Mallet2019} . However this process has been mostly studied for Alfv\'enic turbulence, and it is still unclear how it affects compressive fluctuations. Furthermore, this effect cannot explain the plateau seen in the scale dependent kurtosis between ion and electron scales \cite{Mallet2019} seen in Fig \ref{fig:fig4} \cite[and also in][]{Chen2014}. Coherent wave-wave interactions can also generate coherent structures \cite[e.g.][]{Howes2016} increasing the overall intermittency. Should structures be generated from waves they are still different entities, with very different Fourier signatures. Should there be no randomization, waves would interact and interfere leading to only large scale structures being present.  It seems reasonable to assume that some degree of randomization is required in the cascade. The interaction of waves that are incoherent with respect to any coherent structure could then cause a reduction in global statistics such as the kurtosis.

  Typically in the sub-ion range intermittency destroying waves may be considered to come from an instability or another process external to the turbulent cascade itself. However, following the suggestion that the majority of intervals of solar wind plasma are unstable \cite{Klein2018} it is not clear whether we can disentangle these two ideas. For example, a current sheet could form naturally in the turbulence, become unstable and excite waves. In this case, we cannot truly separate the turbulent cascade from local kinetic plasma processes. Although in the observations of \cite{Klein2018} it is not clear whether the instabilities observed are dynamically relevant as their inverse growth rates in the majority of unstable time intervals are large compared to the turbulence' timescales. There is no indication in this interval of an instability being present, however the ion temperature measurements from MMS are not sufficiently accurate to rule this out. It is unclear where the Ion Bernstein waves originate from, one possibility is that they also cascade from large to small scales but the powers are very small at large scales, and they only become important far into the sub-ion range. The linear damping of both KAWs and IBWs increases with $\beta$. Therefore, we expect that these waves play a smaller role in the magnetosheath and thus allow for a larger overall kurtosis as observed \cite{Chhiber2018}. Curiously, this might affect plasmas such as the magnetosheath with higher $\beta$ and where waves such as those observed here cannot exist may still satisfy Taylor's hypothesis while lower $\beta$ plasmas might not. Our dispersion relation result is also consistent with the interval of \cite{Gershman2018} where a low $\beta$ interval of magnetosheath was investigated. Finally, we remark that this study is limited to a limited scale range. To overcome this limitation we have compared with Cluster data to investigate larger scales. Ideally, all of the relevant scales should be sampled \emph{simultaneously} which can only be achieved from a measurement perspective by surpassing four-point measurements such as the proposed Helioswarm concept \cite{Klein2019}.

\begin{acknowledgments}

The datasets analyzed for this study can be found in the MMS science data archive \url{https://lasp.colorado.edu/mms/sdc/public/}. The calibrated spacecraft potential data are available at \url{https://www.iwf.oeaw.ac.at/en/research/research-groups/space-plasma-physics/sc-plasma-interaction/mmsaspoc-data-analysis/} Analysis of the spacecraft potential data at IWF is supported by Austrian FFG projects ASAP15/873685. D.V. is supported by the STFC Ernest Rutherford Fellowship ST/P003826/1 and STFC Consolidated Grant ST/S000240/1. R.N. was supported by Austrian FWF projects I2016-N20. Z. V. was supported by the Austrian FWF projects P28764-N27.

\end{acknowledgments}


%


\end{document}